\documentclass[12pt]{report}

\usepackage[dvips]{graphicx}
\usepackage{amssymb}
\usepackage{amsmath}

\makeatletter
\renewcommand{\@oddhead}{\hfil \thepage}
\renewcommand{\@oddfoot}{}

\makeatother  
\begin{document}
\begin{center}
\textbf{\large {Evolution of the Quantum Friedmann Universe
Featuring Radiation}} \footnote{$\ $ Published in \textbf{Physics
of Atomic Nuclei}, Vol. 62, No. 4, 1999, pp. 708-714. Translated
from \textbf{Yadernaya Fizika}, Vol. 62, No.4, 1999, pp. 758-764.
}
\end{center}
\begin{center}
\textbf{V. V. Kuzmichev} \footnote{$\ $ e-mail:
vvkuzmichev@yahoo.com; specrada@bitp.kiev.ua}
\end{center}
\begin{center}
\textit{Bogolyubov Institute for Theoretical Physics, National
Academy of Sciences of Ukraine, Metrolohichna St. 14b, Kiev, 03143
Ukraine}
\end{center}

\textbf{Abstract:} The classical and quantum models of the
Friedmann universe originally filled with a scalar field and
radiation have been studied. The radiation has been used to
specify a reference frame that makes it possible to remove
ambiguities in choosing the time coordinate. Solutions to the
Einstein and Schr\"{o}dinger equations have been studied under the
assumption that the rate of scalar-field variation is much less
than the rate of universe expansion (contraction). It has been
shown that, under certain conditions, the quantum universe can be
in quasistationary states. The probability that the universe goes
over to states with large quantum numbers owing to the interaction
of the scalar and gravitational fields is nonzero. It has been
shown that, in the lowest state, the scale factor is on order of
the Planck length. The matter- and radiation-energy densities in
the Planck era have been computed. The possible scenarios of
Universe evolution are discussed.

\begin{center}
1. INTRODUCTION
\end{center}

That quantum gravity theory cannot rely on experimental data [1]
adds importance to exactly soluble cosmological models. However,
the application of basic ideas underlying quantum theory to a
system of gravitational and matter fields runs into difficulties
of a fundamental character, which do not depend on the choice of a
specific model. By way of example, we will consider a homogeneous,
isotropic, and closed universe characterized by the
Friedmann-Robertson-Walker metric; that is,
\begin{eqnarray}
   ds^{2} = a^{2}(\eta )\,[\, N^{2}(\eta )\,d\eta ^{2} -
             d\Omega ^{2}\,],
\label{1}
\end{eqnarray}
Here, $N(\eta)$ is a function that specifies the time-reference
scale; $a(\eta)$ is a scale factor; $d \Omega ^{2}$ is an interval
element on a unit 3-sphere; and $\eta$ is the parameter that is
related to the synchronous proper time $t$ by the differential
equation $dt = N\,a\,d\eta$. Considering that scalar fields play a
fundamental role both in quantum field theory and in the cosmology
of the early Universe [2, 3], we assume that, originally, the
Universe was filled with matter in the form of a uniform scalar
field $\phi$. If the field $\phi$ varies slowly in the early
Universe, its potential $V(\phi)$ specifies the vacuum-energy
density (cosmological term) and ensures Hubble expansion.
Restricting our analysis to the case of minimal coupling between
geometry and the scalar field, we represent the action functional
in the conventional form
\begin{eqnarray}
  S  = \int d\eta \,\left[\pi _{a}\,a' + \pi _{\phi }\,
  \phi ' -  H \right],
\label{2}
\end{eqnarray}
where a prime denotes differentiation with respect to $d/d\eta$;
$\pi _{a}$ and $\pi _{\phi}$ are the momenta canonically conjugate
with the variables $a$ and $\phi$, respectively; and $H$ is the
Hamiltonian given by
\begin{equation}
    H = \frac{1}{2}\,N\,\left[ -\,\pi _{a}^{2}
       + \frac{2}{a^{2}}\,\pi _{\phi }^{2} - a^{2}
       + a^{4}\,V(\phi ) \right] \equiv N\, {\cal R}.
\label{3}
\end{equation}
Here, the variables $a$ and $\phi$ are taken, respectively, in
units of the length $l = \sqrt{2\,G / 3\,\pi }$ and in units of
$\tilde \phi = \sqrt{3 / 8\,\pi \,G}$. The function $N$ plays the
role of a Lagrange multiplier, and the variation $\delta S/\delta
N$ leads to the constraint equation ${\cal R} = 0$. The structure
of the constraint is such that true dynamical degrees of freedom
cannot be singed out explicitly. This creates problems in the
interpretation of quantum geometrodynamics [4—6]. It is commonly
thought that the main reason behind such difficulties is that
there is no natural way to define a spacetime event in general
covariant theories [7]. In the model being considered, the above
difficulties are reflected in that the choice of the time variable
is ambiguous.

For the choice of the time coordinate to be unambiguous, the model
must be supplemented with a coordinate condition. When the
coordinate condition is added to the field equations, their
solution can be found for a fixed time variable. However, this
method of removing ambiguities in specifying the time variable
does not solve the problem of a quantum description, because
undesirable consequences of this ambiguity in eventual equations
cannot be avoided in this way.

In this study, we propose specifying a reference frame with the
aid of an additional matter source. This method does not come into
conflict with the adopted ideas of the early Universe [2, 3]. At
the same time, it enables us to study the evolution of the
Universe not only in the semiclassical approximation but also at a
purely quantum level.

\begin{center}
2. CLASSICAL DESCRIPTION \\ \textit{2.1. Fundamentals of the
Model}
\end{center}

The ambiguity associated with choosing the time coordinate in (1)
will be removed with the aid of a coordinate condition imposed
prior to varying the action functional, but its coordinate
invariance will be restored [7, 8]. We will choose the coordinate
condition in the form $T' = N$, where $T$ is the privileged time
coordinate, and include it in the action functional with the aid
of a Lagrange multiplier $P$; that is,
\begin{equation}
  S = \int \! d\eta\, \left[\,\pi _{a}\,a' + \pi _{\phi }\,\phi '
     + P\,T' - {\cal H}\,\right],
\label{4}
\end{equation}
where
\begin{equation}
 {\cal H} = N\,[\,P + {\cal R}\,]
\label{5}
\end{equation}
is the new Hamiltonian. The constraint equation reduces to the
form
\begin{equation}
   P + {\cal R} = 0.
\label{6}
\end{equation}
Integrating the canonical equation $P' = [P, {\cal H}] = 0$, we
immediately obtain $P = E$, where $E$ is a constant. The full set
of equations for the model in question becomes
\begin{equation}
  \dot{a}^{2} - \frac{a^{2}}{2}\,\dot{\phi }^{2} + U = E,
\label{7}
\end{equation}
\begin{equation}
  \ddot{\phi } + 2\,\frac{\dot{a}}{a}\,\dot{\phi } +
   a^{2}\,\frac{dV}{d\phi } = 0,
\label{8}
\end{equation}
where overdots denote differentiation with respect to$T$ and $U
\equiv a^{2} - à^{4}\,V(\phi )$. Equation (7) represents the
Einstein equation for the $\left(^{0}_{0}\right)$ component, while
equation (8) is the equation of motion for the field $\phi $. A
modification to the Einstein equations that is associated with
including the coordinate condition in the action functional is
that, on the right-hand side, there additionally arises an
energy-momentum tensor $ \tilde T^{0}_{0} = E / a^{4},\
 \tilde T^{1}_{1} = \tilde T^{2}_{2} = \tilde T^{3}_{3} =
 -\,E / 3 a^{4} ,\ \tilde T^{\alpha }_{\beta } = 0
  \ \ \mbox{for} \ \  \alpha \neq \beta $ that can be interpreted
as the energy-momentum tensor of radiation [9]. The choice of
radiation as the matter reference frame is natural for the case in
which relativistic matter (electromagnetic radiation, neutrino
radiation, etc.) is dominant at the early stage of Universe
evolution. If our Universe were described by the model specified
by equation (4), it would be possible to relate the above
radiation at the present era to cosmic microwave background
radiation.

\begin{center}
\textit{2.2. Solving the Einstein Equations}
\end{center}

A feature peculiar to the model in question is that it involves a
barrier in the variable $a$. This barrier, described by the
function $U$, is formed by the interaction of the scalar and
gravitational fields. It exists for any form of the scalar-field
potential $V(\phi )$ and becomes impenetrable on the side of small
$a$ in the limit $V \rightarrow 0$. In just the same way as in
inflation models (see [2, 3]), we assume that the rate at which
the scalar field $\phi $ changes is much smaller than the rate of
universe evolution, $|\dot{a} / a| \gg |\dot{\phi }|$. In this
case, equation (7) can be integrated in a general form. Presented
immediately below are explicit solutions in the regions $a \leq
a_{1}$ and $a \geq a_{2}$, where they are assumed to satisfy the
boundary conditions $a(0) = 0$ and $a(t_{in}) = a_{2}$,
respectively; here, $a_{1}$ and $a_{2}$ are the turning points
($a_{1} < a_{2}$) specified by the condition $U = \epsilon $, and
$t_{in}$ is the initial instant of time in the second region. We
have
\begin{equation}
   a(t) = \left[\frac{1}{2 V}\,\left(1 - \mbox{cosh}\,2 \sqrt{V} t \right)
   + \sqrt{\frac{\epsilon }{V}}\,\mbox{sinh}\,2 \sqrt{V} t \right]^{1/2}
\label{9}
\end{equation}
for $a \leq a_{1}$ and
\begin{equation}
   a(t) = \left\{ \frac{1}{2 V}\,\left[ 1 + \sqrt{1 - 4 V \epsilon }\,
          \mbox{cosh}\,2 \sqrt{V}( t - t_{in} ) \right] \right\}^{1/2}
\label{10}
\end{equation}
for $a \geq a_{2}$. The quantities $\epsilon $ and $U$ depend
parametrically on $\phi $. In the zero-order approximation, the
former is given by $\epsilon  = E$. The above solution to equation
(7) can be refined by taking into account a slow variation of the
field $\phi $ with the aid of the equation
\begin{equation}
    -\,\frac{a^{2}}{2}\,\dot{\phi }^{2} + \epsilon  = E,
\label{11}
\end{equation}
where $\epsilon (\phi )$ stands for a potential term, which is
bounded by the inequality $E \leq \epsilon \leq 1/4 V$. The case
of  $\epsilon > 1/4 V$, which corresponds to an infinite motion,
will not be considered in this study. From equations (8) and (11),
it follows that, in general, a change in the potential $V(\phi )$
entails a change in the quantity $\epsilon (\phi )$.

The solution in (9) describes the universe expanding from the
point of the initial cosmological singularity to the maximum
possible value of $a_{1}$ achieved at the instant $t_{m} =
\frac{1}{4\,\sqrt{V}}\,\ln \left(\frac{1 + 2\,
   \sqrt{V \epsilon }}{1 - 2\,\sqrt{V \epsilon }} \right)$; after
that, the expansion gives way to contraction, and the universe
collapses by the instant $t_{c} = 2\ t_{m}$. For $2\,\sqrt{V}\,t
\ll 1$, the solution in (9) takes the form
\begin{equation}
     a(t) \simeq \left[2\,\sqrt{\epsilon }\,t \right]^{1/2}.
\label{12}
\end{equation}
It is independent of $V$ and describes the evolution of the
universe that is dominated by radiation [9] and which expands in
the de Sitter mode from the point $a = a_{2}$. In the extreme case
of $\epsilon = 0$, where there is no radiation, the region $a \leq
a_{1}$ contracts to the point $a = 0$, and the expansion can
proceed only from the point $a = a_{2}$. Since the region $a <
a_{2}$ cannot be treated in terms of classical theory, it is
assumed that the classical spacetime with $a = a_{2}$ is formed as
the result of a tunnel transition from "nothing" taken to mean
some quantum state of the protouniverse (see, for example, [2,
3,10,11]). If, originally, the universe was filled not only with
matter but also with radiation, it can undergo evolution in the
region $a \leq a_{1}$ as well. In the general theory of
relativity, the solutions in (9) and (10) describe two independent
scenarios of the evolution. The inclusion of the mechanism of
quantum tunneling through the barrier $U$ requires a joint
analysis of these scenarios. It is then legitimate to consider the
probabilities of finding the universe in each of the classically
accessible regions.

The evolution of the universe depends on the initial distribution
of the classical field $\phi $ and its subsequent behavior as a
function of time. The chaotic-inflation scenario [3], which is
realized in the region $a > a_{2}$, is described by equations (7)
and (8) as applied to the case specified by the inequalities
$(\frac{d \ln V}{d \phi })^{2} \ll 1, \ V \gg | \frac{1}{a^{2}} -
\frac{\epsilon }{a^{4}}|, \ \mbox{and} \ \frac{1}{a^{2}} |
\ddot{\phi } | \ll | \frac{d V}{d \phi }|$. In the model where the
scalar-field potential $V$ is taken to be proportional to $\phi
^{n}$, the chaotic-inflation process proceeds between scalar-field
values greatly exceeding a level of $\frac{n}{3 \sqrt{2}}$
(initial stage) and those achieving this level (final stage). In
this approach, radiation has virtually no effect on the degree of
inflation, and the scalar field represents the field of an
inflaton. The de Sitter regime of inflation persists as long as
the potential $V(\phi (t))$ varies rather slowly with time. From
equations (8) and (11), it follows that the inequality $\dot V +
\dot \epsilon / a^{4} < 0$ holds in the expanding universe $(\dot
a > 0)$. If the potential $V$ increases with time, the quantity
$\epsilon $ is bound to decrease. But if $V$ decreases, $\epsilon$
can increase, and the rate of this increase is higher for greater
$a$. We will now estimate $\epsilon $ by using the relation
$\epsilon \simeq \tilde T^{0}_{0} a^{4}$. In our Universe, with $a
\sim 10^{28}$ cm, the main contribution to the radiation-energy
density comes from cosmic microwave background radiation with
energy density $\rho _{\gamma }^{0} \sim 10^{- 10}$
GeV/$\mbox{cm}^{3}$. Setting $\tilde T^{0}_{0} = \rho _{\gamma
}^{0}$, we find that, at the present era, the result is $\epsilon
= \epsilon _{\gamma } \sim 10^{117}$. In the early Universe, the
scale parameter is $a \sim 10^{- 33}$ cm, while the energy density
$\tilde T^{0}_{0} $ is on the order of the Planck value. On this
basis, it can be found that $\epsilon \sim 1$ corresponds to that
era. It follows that $\epsilon $ increased in the evolution
process. This increase can be explained by a considerable
redistribution of energy between the scalar field and radiation at
the initial stage of Universe existence. Quantum theory is able to
account for this phenomenon in a natural way (see below). In the
region $a > a_{2}$, the possible variation of $\epsilon $ with
time does not affect the quasiexponential expansion of the
universe, because the inflation stage terminates in a rather short
time interval of $t \sim 10^{- 37}$ s [3], and the evolution
process is then determined by other factors (particle production,
heating, etc.). In the region $a < a_{1}$, the role of the
increase in $\epsilon $ with decreasing $V$ may prove substantial.
In principle, the dependence of $\epsilon $ on $\phi (t)$ makes it
possible to provide the missing power in the $t$ dependence of $a$
and to solve the problem of the Universe size. In order to
demonstrate this explicitly, we assume that, up to the present
time $t_{0}$, the Universe has expanded according to the law
specified by (12) [2]. We then have $a(t_{0}) / a(t_{p}) =
\left(\sqrt{\epsilon _{0} / \epsilon _{p}} \ t_{0}/ t_{p}
\right)^{1/2} $, where $\epsilon _{0} = \epsilon (\phi (t_{0}))$,
$\epsilon _{p} = \epsilon (\phi (t_{p}))$, and $t_{p}$ is the
Planck time. The ratio of $\epsilon _{0}$ and $\epsilon _{p}$ can
be estimated as $\epsilon _{0}/\epsilon _{p} \sim V_{p}/V_{0}$,
where $V_{p} = V (\phi (t_{p}))$ and $V_{0} = V (\phi (t_{0}))$.
Assuming that, in the Planck era, $V_{p}$ is on the order of the
Plank energy density and that $V_{0}$ is on the order of the mean
matter-energy density at the present era, $\rho _{0} = 10^{-5}
\mbox{GeV}/\mbox{cm}^{3}$, we find that the value of $a(t_{0})
\sim 10^{28}$ cm corresponds to $a(t_{p}) \sim 10^{-33}$ cm.

\begin{center}
3. QUANTIZATION \\ \textit{3.1. Schr\"{o}dinger Equation}
\end{center}

In quantum theory, the constraint equation (6) comes to be a
constraint on the wave function that describes the universe filled
with a scalar field and radiation. Replacing the canonically
conjugate variables involved in equation (7) by the operators $
\hat{a} = a \times, \ \hat{\pi _{a}} = -\,i\,\partial _{a},
           \ \hat{\phi } = \phi \times,
           \ \hat{\pi _{\phi }} = -\,i\,\partial _{\phi },
           \ \mbox{and} \hat{P} = -\,i\,\partial _{T}$, we find that the
state vector $\langle a, \phi | \Psi (T) \rangle $ satisfies the
equation $\langle a, \phi | \Psi (T) \rangle $
\begin{equation}
   2\,i\, \partial _{T}| \Psi (T) \rangle = \left[ \partial _{a}^{2} -
  \frac{2}{a^{2}}\,\partial _{\phi }^{2} - U \right] | \Psi (T) \rangle.
\label{13}
\end{equation}
where the order parameter is assumed to be zero [5, 10-13].
Equation (13) represents an analog of the Schr\"{o}dinger equation
with a Hamiltonian independent of the time variable $T$. The
momentum $\hat{P}$ associated with radiation appears linearly in
equation (13). We can introduce a positive definite scalar product
$\langle \Psi | \Psi \rangle < \infty $ and specify the norm of a
state [8, 11]. This makes it possible to define a Hilbert space of
physical states and to construct quantum mechanics for the
universe model being considered.

A partial solution to equation (13) has the form
\begin{equation}
  | \Psi (T) \rangle = | \psi \rangle
  \exp \left\{\frac{i}{2}\,E \left( T - T_{0} \right) \right\},
\label{14}
\end{equation}
where the state $\psi $ satisfies the time-independent equation
\begin{equation}
 \left( -\,\partial _{a}^{2} + \frac{2}{a^{2}}\,\partial _{\phi }^{2} +
             U - E  \right) | \psi \rangle = 0.
\label{15}
\end{equation}
The quantity $E$ is arbitrary in the general theory of relativity,
but, in quantum theory, it is quantized in accordance with
solutions to equation (15).

\begin{center}
\textit{3.2. Quasistationary States}
\end{center}

In considering the quantum case, we assume that, at the initial
stage, the motions occurring in the system under study can be
separated into two types: a slow variation of the scalar field, in
which case the operator $(2 / a^{2}) \partial _{\phi }^{2}$ can be
treated as a perturbation, and fast changes in geometry. This
assumption is a quantum analog of the adiabaticity hypothesis,
which leads, in the zero-order approximation, to the solutions
given by (9) and (10). In quantum theory, the problem being
considered reduces to solving the equation
\begin{equation}
 \left[ \partial _{a}^{2} - U + \epsilon _{n}(\phi ) \right]
        |\varphi _{n} \rangle = 0.
\label{16}
\end{equation}
The wave functions $\varphi _{n}$ and the eigenvalues $\epsilon
_{n}$, which depend on $\phi $ parametrically, describe the
evolution of the universe for very slow variations of the
potential $V$ associated with the field $\phi $ (more
specifically, under the condition $|d \ln V / d \phi | \ll 1$). In
order to take into account the variations of the field $\phi $, we
can represent $\psi $ as an expansion in terms of the states
$\varphi _{n}$ and integrate then equation (15). The quantum
number $n$ of the system unperturbed by the operator $(2 / a^{2})
\partial _{\phi }^{2} $ will be a good quantum number for
the universe in the perturbed state $\psi $ as well.

We will further consider solutions to equation (16), allowing for
the possible boundary conditions. For $\epsilon _{n} \leq 1/4 V$
the classically accessible regions $a \leq a_{1}$ and $a \geq
a_{2}$ are bounded by the turning points $a_{1}$ and $a_{2}$,
which are now dependent on the state of the quantum system. In the
region $a > a_{2}$, a general solution has the form of a
superposition of converging and diverging waves. In the
Wentzel-Kramers-Brillouin (WKB) approximation, we can write
\begin{eqnarray}
  \varphi _{n}  =  \frac{1}{( \epsilon _{n} - U )^{1/4}}\,
      \left\{ C_{1}\,\mbox{e}^{\, i \int \limits_{a_{2}}^{a}
             \sqrt{ \epsilon _{n} - U }\,da - \frac{i \pi }{4}} +
                  C_{2}\,\mbox{e}^{ -\,i\int \limits_{a_{2}}^{a}
       \sqrt{ \epsilon _{n} - U }\,da + \frac{i \pi }{4} } \right\},
\label{17}
\end{eqnarray}
where $C_{1}$ is the amplitude of an "incident" wave describing
the universe whose scale factor decreases, while $C_{2}$ is the
amplitude of the wave "traveling" toward greater values of $a$ and
describing the expanding universe. For the extreme case of
$\epsilon _{n} = 0$, which corresponds to the radiation-free
universe with an undetermined time variable, the wave function in
the form (17) was studied by many authors (see, for example, [3,
5, 10-13]). It coincides with the Vilenkin wave function [10] at
$C_{1} = 0$ and generalizes the Hartle-Hawking wave function [11]
to the case of $\epsilon _{n} \neq 0$.

If we consider a universe formed at an instant separated by a
comparatively large time interval [$-\,\infty < ( T - T_{0} ) \leq
0$] from the commencement of observation, the boundary condition
$C_{1} = 0$ will imply that the diverging wave corresponding to a
quasistationary state [14] is singled out from the superposition
in (17). No situation that is physically realizable can exactly
correspond to the requirement $C_{1} = 0$ for all instants of time
because, in that case, the process that leads to the formation of
a quasistationary state and which involves converging waves would
be eliminated from the analysis. According to the general concepts
of quantum theory [14, 15], a quasistationary state can be
implemented approximately by requiring that the region where the
asymptotic form (17) with $C_{1} = 0$ is realized be bounded by
the condition $a \leq a_{max}$, in which cases $\varphi _{n}$ is
set to zero for $a > a_{max}$, $a_{max} \sim \sqrt{\epsilon _{n}}
T$ being some boundary value of the scale factor. That the
instants of time that satisfy the conditions $T < T_{0}$ and $T >
a_{max}/\sqrt{\epsilon _{n}}$ are excluded from the analysis is
physically justified because this makes it possible to avoid
speculations about the properties of the scalar field and
radiation in the Universe at times that are practically
inaccessible to observation and for which there are no reliable
hints from high-energy physics.

The possibility of quantum tunneling through the region $a_{1}
\leq a \leq a_{2}$ of the potential barrier results in that
stationary states cannot be realized in the region $a \leq a_{1}$.
If, however, the potential $V(\phi )$ is sufficiently small,
quasistationary states whose lifetime is much greater than the
Planck time can exist in the region $a \leq a_{1}$. The
probability $\Gamma _{n}$ of the decay of the universe occurring
in a given quasistationary state $\varphi _{n}$ can be found by
requiring that the wave function in (17) satisfy the radiation
condition $C_{1} = 0$, which selects the discrete complex values
$\tilde \epsilon _{n} = \epsilon _{n} + i \Gamma _{n}$[14].

We further impose the boundary condition $ \varphi _{n} \left|_{a
= 0} = 0 \right. $ on the wave function $\varphi _{n}$. For the
case of $\Gamma _{n} \ll \epsilon _{n}$, we then obtain
\begin{eqnarray}
   \Gamma _{n} = 2\,\left[\int \limits_{0}^{a_{1}}
               \frac{da}{\sqrt{\epsilon _{n} - U}} \right]^{- 1}
               \exp\left\{ -\,2 \int \limits_{a_{1}}^{a_{2}}
               \sqrt{U - \epsilon _{n}}\, da  \right\},
\label{18}
\end{eqnarray}
where $\epsilon _{n}$ is determined from the equation
\begin{equation}
     \int \limits_{0}^{a_{1}} \sqrt{\epsilon _{n} - U}\, da =
              \pi \left( n + \frac{3}{4} \right).
\label{19}
\end{equation}
In the extreme case of $V = 0$, equation (19) can easily be
integrated, which yields $\epsilon _{n} |_{V = 0} \equiv \epsilon
_{n}^{(0)} = 4 n + 3$; that is, we can see that, for all values of
$n$, $\epsilon _{n}^{(0)}$ coincides with the energy of an
isotropic oscillator with zero orbital angular momentum [16, 17].
According to (19), the first level (that at $\epsilon _{0} \sim
3$) emerges at $V \sim 0.08$.

\begin{table}
\caption{Values of $\epsilon _{n}$ as computed in the WKB
approximation on the basis of (19) and within perturbation theory
(PT) for various values of the potential $V$ (here, $U_{max} =
1/4V$ is the barrier height; $\Delta a = a_{2} - a_{1}$ 1s the
barrier width; and $\Sigma $ is the number of levels at given
$V$), along with the decay-probability values $\Gamma _{n}$ as
given by (18) under the same conditions}

\begin{center}
\begin{tabular}{c|c|c|c|c|c|c|c}    \hline
$n$ & $V$    & $\epsilon _{n}$(WKB)   & $\epsilon _{n}$(PT)    &
$\Gamma _{n}$        & $U_{max}$ & $\Delta a$ & $\Sigma $ \\
\hline 0 & 0.08 & 2.62                 & 2.63                 &
0.31 & 3.125 & 1.03 & 1       \\
  & 0.05 & 2.79                 & 2.79                 & 0.006              & 5       & 2.25     & 1       \\
  & 0.03 & 2.89                 & 2.88                 & 2 $\times 10^{- 6}$  & 8.33    & 3.70     & 2       \\
  & 0.02 & 2.93                 & 2.92                 & 7 $\times 10^{- 11}$ & 12.5    & 5.08     & 3       \\
  & 0.01 & 2.97                 & 2.96                 & $10^{- 24}$          & 25      & 8.05     & 6       \\ \hline
1 & 0.03 & 6.34                 & 6.35                 & 0.01
& 8.33    & 2.06     & 2       \\
  & 0.02 & 6.59                 & 6.59                 & $10^{- 6}$           & 12.5    & 3.70     & 3       \\
  & 0.01 & 6.80                 & 6.80                 & $10^{- 19}$          & 25      & 6.92     & 6       \\  \hline
2 & 0.02 & 9.88                 & 9.94                 & 0.003
& 12.5    & 2.34     & 3       \\
  & 0.01 & 10.51                & 10.51                & $10^{- 15}$          & 25      & 5.93     & 6       \\  \hline
\end{tabular}
\end{center}
\end{table}

At small $V$, equation (19) leads to $\epsilon _{n}$ values
coincident with those that are obtained directly from equation
(16) by perturbation theory in $a^{4} V(\phi )$. The table
displays $\epsilon _{n}$ values calculated by perturbation theory
and in the WKB approximation [that is, with the aid of equation
(19)]. Also presented in this table are the decay-probability
values $\Gamma _{n}$ computed for various potentials $V$. Since
$\Gamma _{n} \ll \mbox{Re}\,\tilde{\epsilon }_{n}$, the
decay-probability values $\Gamma _{n}$ found on the basis of (18)
are expected to be close to true values for small $n$ as well.

We note that the smaller the value of $a^{4} V$ at given $\epsilon
_{n}$, the higher and the broader is the potential barrier $U$
and, hence, the smaller is the decay probability $\Gamma _{n}$. If
some state $\varphi _{n}$ is characterized by a small value of
$\Gamma _{n}$, the possibility that this state decays can be
disregarded over the decay time $\tau = 1 / \Gamma _{n}$, so that
this state can be considered to be stationary in this limit. This
corresponds to defining a quasistationary state as that which
takes the place of a stationary state when the probability of its
decay becomes nonzero [14].

In describing the universe on the basis of equation (15), the
process of universe production from "nothing" in the
radiation-free model ($E = 0$) [2, 10, 12, 13] is replaced by
quantum tunneling from a quasistationary state with a definite
(complex) value of $E$. It can easily be seen that the solution
given by (17) describes the de Sitter regime of expansion
according to (10). In order to demonstrate this explicitly, we
note that, in the WKB approximation,   we  have $-\,i\,\partial
_{a} \varphi _{n} \approx -\,\sqrt{\epsilon _{n} - U}\,\varphi
_{n}$; that is, the classical momentum is given by $\pi _{a} =
-\,\dot a = -\,\sqrt{\epsilon _{n} - U}$, whence we obtain
equation (7) in approximation $|\dot a / a | \gg |\dot \phi |$.

In quantum models not featuring radiation, the DeWitt-Wheeler
equation [5, 18] determines the time-independent wave function of
the universe. This leads to well-known difficulties in
interpreting this function and in comparing results obtained on
its basis with the observed evolution of our Universe [3, 6]. If,
however, the case of $E = 0$ is considered as the limit to which
the model with a privileged reference frame reduces when $E
\rightarrow 0$, we can also speak about the time evolution of the
Universe free from radiation [19].

\begin{center}
\textit{3.3. Dynamics in the Prebarrier Region}
\end{center}

By studying inflationary scenarios of the evolution of a universe
filled with a scalar field, it was revealed that a "realistic"
potential $V$ must decrease with time [10]. With decreasing $V$,
the number of quantum states in which the universe can occur
increases, while the decay probabilities decrease sharply (see
table). The first instants of the existence of the universe are
especially favorable for its tunneling through the potential
barrier $U$.

A quasistationary state $\varphi _{n}$ takes the place of the
stationary state whose wave function in the region $a < a_{1}$; is
close to the wave function $| n \rangle $ of the state unperturbed
by the interaction $a^{4} V$. In the approximation of a slowly
varying field $\phi $, transitions in the system being studied can
be considered as those that occur between the states $| n \rangle
$ and which are induced by the interaction $a^{4} V$. Since this
interaction modifies the physical properties of the system, a
finite number of its levels and their nonzero widths must be taken
into account in calculating the probabilities $W_{n m}$ of the
$m(T_{0}) \rightarrow n(T)$ transitions. As a result, we arrive at
\begin{equation}
   W_{n m} \approx \left| \langle n | {\cal U}_{I}(T,T_{0})
            | m \rangle \right|^{2}\,
           \exp\left\{ -\,\Gamma _{n} \Delta T \right\},
\label{20}
\end{equation}
where $\Delta T = T - T_{0}$ and ${\cal U}_{I}$ is the evolution
operator in the interaction representation [20]. Adiabaticity in
the field $\phi $ enables us to consider specific transitions in
the time interval $\Delta T$ that correspond to a given value of
$V(\phi )$.

The figure displays the total probability of universe decay, $
W_{dec} = 1 - \left( W_{0 0} + W_{1 0} \right)$, and the quantity
$W_{1 0}$ as calculated at $V = 0.03$, in which case there are
only two levels in the system. It can be seen that, over the time
interval $\Delta T \lesssim 50$, the transitions in the system
predominate and only for $\Delta T \sim 100$ the probability that
the universe tunnels through the barrier becomes commensurate with
the probability that it undergoes the $0 \rightarrow 1$ transition
in the prebarrier region.

\begin{figure}
\begin{center}
\includegraphics[scale=1.15]{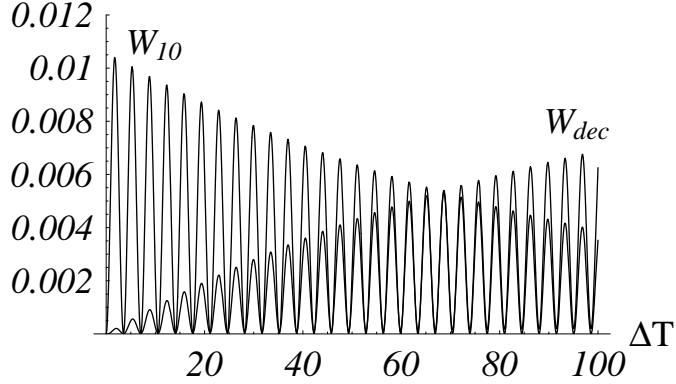}
\end{center}
\vspace{-2.cm} \caption{Probabilities $W_{10}$ and $W_{dec}$
versus the time interval $\Delta T = T - T_{0}$ at the parameter
values of $V = 0.03,\ \epsilon _{0}^{(0)}= 3,\ \epsilon _{1}^{(0)}
= 7,\ \Gamma _{0} = 2 \times 10^{- 6},\ \mbox{and} \ \Gamma _{1} =
10^{- 2}$.}
\end{figure}

Since the rate at which the level width $\Gamma _{n}$ tends to
zero is greater than the rate at which the potential decreases,
the reduction of $V$ with time results in that transitions become
much more probable than tunnel decays, in which case the former
fully determine the evolution of the quantum universe in the
prebarrier region. If the universe has not tunneled through the
barrier before the potential $V$ of the field $\phi $ decreases to
a value less than $0.01$, a sufficiently large number of levels
such that the probabilities of decays from them can be neglected
are formed in it. Assuming that the amplitudes of transitions over
the time interval $\Delta T $ are small, $\left| V\,\langle n |
a^{4} | m \rangle \right|\,\Delta T \ll 1$, we then find that
\begin{eqnarray}
   \frac{W_{n + 1, n}}{W_{n - 1, n}} > 1, \quad
   \frac{W_{n + 1, n}}{W_{n + 2, n}} > 1, \quad
   \frac{W_{n + 2, n}}{W_{n - 2, n}} > 1, \quad
   \frac{W_{n - 1, n}}{W_{n + 2, n}} > 1,
\label{21}
\end{eqnarray}
that is, the $ n \rightarrow n + 1 $ transition is more probable
than the $ n \rightarrow n - 1$ and $n \rightarrow n + 2$
transitions. This means that the quantum universe can undergo
transitions to ever higher levels with a nonzero probability. It
is well known that the oscillator amplitude is quantized according
to the condition $\bar a \sim \sqrt{n}$; therefore, it can be
concluded that the characteristic size $\bar a $ of the universe
that did not undergo a tunnel transition increases as it is
excited to higher levels.

\begin{center}
\textit{3.4. Parameters of the Early Universe}
\end{center}

In the adiabatic approximation, the expectation value $\bar a $
for the universe occurring in the lowest state $\varphi _{0}$ is
given by
\begin{equation}
   \bar a \approx \langle \varphi _{0} | a | \varphi _{0} \rangle
   = \frac{2}{\sqrt{\pi }}\,\left[ 1 + \frac{21}{16}\,V +
            O\left( V^{2} \right) \right].
\label{22}
\end{equation}
whence it follows that $\bar a \approx 0.9 \times 10^{-
33}\,\mbox{cm}$ for $0 < V < 0.08$. The value $\bar a$ determines
the mean amplitude of oscillations of the classical universe
filled with matter and radiation, thereby specifying its actual
linear dimension. The maximal proper distance in a closed universe
can be estimated at $ d \approx \pi \,\bar a \approx 3 \times
10^{- 33}\,\mbox{cm}$; that is, the universe in the lowest state
has a proper dimension on the order of the Planck length [19]. The
presence of the minimal length removes the problem of the initial
cosmological singularity.

Equations (7) and (8), which are obtained within the general
theory of relativity, also dictate the relationship between the
quantities $a,\ \epsilon $ and $V$. By using the value of
$\epsilon _{0} \simeq 2.6$, which we found for $V \simeq 0.08$, we
can estimate the classical turning points at $a_{1} \simeq 1.4
\times 10^{- 33}\mbox{cm}$ and $a_{2} \simeq 2.2 \times 10^{-
33}\mbox{cm}$. The value of $a_{1}$ determines the maximal
dimension of the universe occurring in the lowest state to the
left of the barrier along the $a$ axis, while $a_{2}$
characterizes its initial dimension upon tunneling from this
state. In this era, the matter- and radiation- energy densities
are $ T_{0}^{0} \approx V \approx  1.3 \times 10^{77}\,
\mbox{GeV}/\mbox{fm}^{3},$ and $ \tilde T_{0}^{0} \approx
\frac{\epsilon _{0}}{\bar a^{4}} \approx 1.7 \times
10^{78}\,\mbox{GeV}/\mbox{fm}^{3}$, respectively; that is, we can
see that, according to our model, the energy density in the early
universe is determined primarily by the radiation-energy density.
This result is fully consistent with what is commonly thought
about the properties of the universe for $a \rightarrow 0$ [9].
The total matter-energy density is $\rho \approx 0.64\,m_{p}^{4}$,
where $m_{p}$ is the Planck mass. Thus, we can see that quite
reasonable results are obtained when the parameters $\epsilon $
and $V$ as derived on the basis of quantum theory are used in the
equations of the general theory of relativity.

It is interesting to estimate the quantity $n$ at the $\bar a$
value coincident with the dimension of the presently observed part
of the Universe. From the relation $\bar a \sim \sqrt{n}$ at $\bar
a \sim 10^{28}\,\mbox{cm}$, we obtain $n \sim 10^{122}$, whence we
can see that, if the quantum model being considered is
extrapolated to the observed Universe, it occurs in a highly
excited state. Quantum corrections to the classical equations of
the general theory of relativity are extremely small in this case
(they are of order $\sim 1/n $). The resulting value of $n \sim
10^{122}$ is consistent with the estimates presented in [11, 21]
and is confirmed by rigorous quantum-mechanical calculations
within the radiation-free model that was considered in [19] and
which is justified for the present, large, values of $\bar a$ at
the matter-dominated stage.

\begin{center}
4. CONCLUSION
\end{center}

The presence of radiation in the universe makes it possible to
associate a privileged reference frame with it and to remove
thereby an ambiguity in choosing the time coordinate. This opens
new possibilities both in classical and in quantum cosmology. Upon
performing quantization, there naturally arises the
Schr\"{o}dinger equation (13) with an effective interaction $U$ in
the form of a potential barrier. The evolution of the universe
involves a quantum stage that is realized in the prebarrier region
and which precedes the process of tunneling through the barrier.
The dynamics of this stage is governed by the interaction of the
gravitational and scalar fields. That the system in question
possesses a spectrum of quantum (quasistationary) states and that
transitions can occur between these states enable us to take a
fresh look at the problem of the dimension of the Universe. In the
approach developed here, the universe is characterized by a
minimal length, so that the singularity problem does not arise in
it. The probability for the universe to undergo a tunnel
transition is maximal in the lowest quantum state, where the
energy density and the scale factor are on the same orders of
magnitude as the corresponding Planck values. If a quantum
universe tunnels from higher states, the dimensions of the region
from which tunneling occurs can considerably exceed the Planck
length. The constants $E$ and $V$ appearing in the Einstein
equations are determined by the preceding, quantum stage. The use
of the parameters in the general theory of relativity that were
obtained on the basis of quantum theory leads to conclusions that
are consistent with the currently adopted concepts of the early
Universe and its subsequent evolution.

\begin{center}
REFERENCES
\end{center}

1. Isham, C.J., gr-qc/9510063.

2. Dolgov, A.D., Zeldovich, Ya.B., and Sazhin, M.V., Kosmologiya
rannei vselennoi (Cosmology of the Early Universe), Moscow: Mosk.
Gos. Univ., 1988.

3. Linde, A.D., Elementary Particle Physics and Inflationary
Cosmology, Chur: Harwood, 1990.

4. Arnowitt, R., Deser, S., and Misner, C.W., Gravitation: An
Introduction to Current Research, Witten, L., Ed., New York, 1963.

5. DeWitt, B.S., Phys. Rev., 1967, vol. 160, p. 1113.

6. Ponomarev, V.N., Barvinskii, A.O., and Obukhov, Yu.N.,
Geometrodinamicheskie metody i kalibrovochnyi podkhod k teorii
gravitatsionnykh vzaimodeistvii (Methods of Geometrodynamics and
Gauge Approach in the Theory of Gravitational Interactions),
Moscow: Energoatomizdat, 1985.

7. Kucha\v{r}, Ê., Proc. 4th Canadian Conf. on General Relativity
and Astrophysics, Kunstatter, G., Vincent, D., and Williams, J.,
Eds., Singapore: World Sci, 1992.

8. Kucha\v{r}, K.V. and Torre, C.G., Phys. Rev. D: Part. Fields,
1991, vol. 43, p. 419.

9. Landau, L.D. and Lifshitz, E.M., The Classical Theory of Fields,
Oxford: Pergamon, 1975.

10. Vilenkin, A., Phys. Rev. D: Part. Fields, 1994, vol. 50, p.2581.

11. Hartle, J.B. and Hawking, S.W., Phys. Rev. D: Part. Fields,
1983, vol. 28, p. 2960.

12. Vilenkin, A., Phys. Rev. D: Part. Fields, 1986, vol. 33, p. 3560.

13. Vilenkin, A., Phys. Rev. D: Part. Fields, 1988, vol. 37, p. 888.

14. Baz', A.I., Zel'dovich, Ya.B., and Perelomov, A.M.,
Scattering, Reactions, and Decays in Nonrelativistic Quantum
Mechanics, Jerusalem: Israel Program of Sci. Transl., 1966.

15. Blatt, J.M. and Weisskopf, V.R, Theoretical Nuclear Physics,
New York: Springer-Verlag, 1979.

16. Davydov, A.S., Quantum Mechanics, Oxford: Pergamon, 1976.

17. Fl\"{u}gge, S., Practical Quantum Mechanics, Berlin:
Springer-Verlag, 1971.

18. Wheeler, J.A., Battelle Rencontres, DeWitt, C. and Wheeler,
J.A., Eds., New York: Benjamin, 1968, p. 242.

19. Kuzmichev, V.V, Yad. Fiz., 1997, vol. 60, p. 1707
[Phys. At. Nucl. (Engl. Transl.), vol. 60, p. 1558].

20. Dirac, P.A.M., The Principles of Quantum Mechanics,
Oxford: Clarendon, 1958.

21. Zeldovich, Ya.B. and Novikov, I.D., Relativistic Astrophysics,
vol. 2: The Structure and Evolution of the Universe, Chicago:
Univ. of Chicago Press, 1983.

\end{document}